# Excitation of surface and volume plasmons in metal nanocluster by fast electrons

V. B. Gildenburg[1,2,*], V. A. Kostin[1,2], and I. A. Pavlichenko[1,2]
[1]*University of Nizhny Novgorod, Nizhny Novgorod 603950, Russia*
[2]*Institute of Applied Physics, Russian Academy of Sciences, Nizhny Novgorod 603950, Russia*

Collective multipole oscillations (surface and volume plasmons) excited in a metal cluster by moving electron and corresponding inelastic scattering spectra are studied based on the hydrodynamic approach. Along with the bulk (dielectric) losses traditionally taken into account, the surface and radiative ones are also considered as the physical mechanisms responsible for the plasmon damping. The second and third mechanisms are found to be essential for the surface plasmons (at small or large cluster radii, respectively) and depend very differently on the multipole mode order. The differential equations are obtained which describe the temporal evolution of every particular mode as that one of a linear oscillator excited by the given external force, and the electron energy loss spectra are calculated. The changes in spectrum shape with the impact parameter and with the electron passage time are analyzed; the first of them are found to be in good enough agreement with the data of scanning transmission electron microscopy (STEM) experiments. It is shown that, in the general case, a pronounced contribution to the formation of the loss spectrum is given by the both surface and volume plasmons with low and high multipole indices. In particular, at long electron passage time, the integral (averaged over the impact parameter) loss spectrum which is calculated for the free-electron cluster model contains two main peaks: a broad peak from merging of many high-order multipole resonances of the surface plasmons and a narrower peak of nearly the same height from merged volume plasmons excited by the electrons that travel through the central region of the cluster. Comparatively complex dependences of the calculated excitation coefficients and damping constants of various plasmons on the order of the excited multipole result in wide diversity of possible types of the loss spectrum even for the same cluster material and should be taken into account in interpretation of corresponding electron energy loss spectroscopy (EELS) experiments.

## I. INTRODUCTION

The collective electron excitation (surface and volume plasmons) play an important role in formation of polarization response and absorption spectra for metal nanoclusters irradiated by laser pulses or charged particle beams. The corresponding resonances are the subject of unremitting interest which is stimulated by the development of optoelectronic, computer, and biomedical nanotechnologies [1-3]. The relative roles and intensities of different resonances in observed spectra are determined by the nature of the external excitation and the nanoparticle parameters. In particular, the main role in laser-cluster interaction is played by the resonance of the dipole surface plasmon mode. For a spherical cluster, this is the so-called Mie resonance at the frequency $\omega_p/\sqrt{3}$, where $\omega_p$ is the plasma frequency. The most of performed experimental, theoretical, and applied studies on absorption and scattering spectra of a laser-irradiated cluster were focused on exactly this resonance (and its modifications with account of spatial dispersion, particle nonshpericity, nonlinearity, inhomogeneity, quantum size effect, etc [4-10]). The other plasmon resonances, including the surface higher-order multipoles and all the volume plasmons [4], due to their relatively weak coupling with the external optical field are strongly suppressed by the internal losses (as a rule, associated with electron-phonon interaction and interband transitions) even at minimum possible values of the effective electron collision frequency $\nu \sim 0.01\omega_p$ [5, 6].

The situation is different when the plasmons are excited by moving charged particles: the inelastic energy loss spectra for electrons scattered by a cluster can contain resonances related to excitation of surface high-order multipoles and volume plasmons [11-23]. These spectra being the object of the field developed intensively in the recent decades, namely, the electron energy losses spectroscopy (EELS) [24, 25], have been studied to date in sufficient detail both experimentally (for the spherical clusters of various types and sizes and nonspherical nano-objects [20-23, 26-28] and theoretically (on the basis of quantum and semiclassical models [11-19]). Nevertheless, some important theoretical issues related to the features of excitation and physical mechanisms of plasmon damping require further study even in the simplest case of a spherical cluster (especially, in the case, when the electron passes through its internal regions).

In the simplest models (see, e.g., Refs. [12, 13, 29]), the electron loss spectra are calculated with no allowance for the possibility that the scattered electron may travel inside the cluster: the impact parameter is assumed to exceed the cluster radius in the most semiclassical calculations performed. In fact, this excludes the possibility of allowing for volume plasmons and does not allow one to compare the calculation results with the data of the experiments, in which clusters were bombarded with a thin electron beam at different impact parameters (including those being less than the radius) [20-23]. In the works based on the solution of the quantum-mechanical problem about inelastic scattering of a plane de Broglie wave [14-17] (with homogeneous or Gaussian probability distribution along its front), certainly, no such assumption is made, but specific calculations or discussions of the impact parameter dependencies (which are, evidently, difficult to obtain within the general quantum approach used) are absent as well.

In this work, the semiclassical model for inelastic scattering of fast electrons by a spherical cluster is developed and fits any values of the impact parameters both longer and shorter than the cluster radius. This model, as well as the above-mentioned works [11-13, 29], is based on the use of quantum-mechanical correspondence between the frequency- and energy-dependent loss spectra of the electron decelerated at its classical (weakly perturbed) trajectory by the field of all plasmons it excites. Within the model used, the position, relative height, and width of the resonance peaks in the calculated loss spectra are determined, evidently, by the complex eigenfrequencies and excitation coefficients, which depend for every plasmon on its spatial structure (with electron velocity, impact parameter, and cluster radius given). The plasmon damping constant (width of the resonance line) is here a rather important parameter (but apparently the least accurately known one) which determines the role of a particular plasmon in the total loss spectrum. In the general case, the damping constant is determined (i) by the imaginary part of the complex bulk dielectric permittivity of the cluster substance; (ii) by the so-called electron-surface scattering, i.e., by the collisions of cluster free electrons with cluster boundary; and (iii) by the radiative losses caused by electromagnetic radiation of the particular multipole into the surrounding space. The latter two mechanisms of the losses have been studied previously only for the dipole surface plasmon; and thus for calculation the total loss spectrum being of interest, we perform the required generalization of the known formulas for the damping constants of dipole oscillations to multipoles of arbitrary order. The general formula for the radiation damping constant was obtained just through allowance for additional wave corrections in the known solution of problem on dielectric sphere oscillation. The boundary scattering damping constant was calculated based on a comparatively simple kinetic model which describes the conversion of the oscillatory kinetic energy component into the static (thermal) one during reflection of the free electron in cluster plasma from plasma boundary. Our calculations demonstrate the great variety of possible shapes of loss spectra for the same cluster material and are in good agreement with the experimental data.

The paper is organized as follows. In Section II we describe the spectrum of eigenoscillations of the plasma in a spherical metal cluster, provide formulas for the real eigenfrequencies of the different multipole modes (including as surface as volume plasmons), and calculate their damping constants determined by the internal (bulk and surface) and radiative losses. In Section III based on the found orthogonality relation, we develop the general excitation equations that allow one to describe temporal evolution of plasmons as a system of independent oscillators with given external forces and obtain the formula for the electron energy loss spectrum. In Section IV we present and discuss the loss spectra calculated numerically for different external parameters and various cluster materials and compare the results obtained with the data of experiments and other calculations. In Section V, the final conclusions are formulated. In Appendix, the orthogonality relation we used in Section 3 is derived.

## II. SURFACE AND VOLUME PLASMONS IN A SPHERICAL NANOCLUSTER

Being accounted within the hydrodynamic approach [30], the dynamic polarization nonlocality in metal degenerate plasma originates in the following equation relating the complex vector amplitudes of alternating electric field $\mathbf{E}\exp(-i\omega t)$ and displacement $\mathbf{D}\exp(-i\omega t)$:

$$\mathbf{D} = \varepsilon \mathbf{E} + \frac{3}{5}\varepsilon_\infty r_F^2 (\varepsilon_\infty - \varepsilon)\nabla(\nabla \mathbf{E} - \nabla \mathbf{D}). \tag{1}$$

Here, $r_F = V_F/\omega_p$ is the Thomas–Fermi length; $V_F$ is the Fermi velocity; $\omega_p = \sqrt{4\pi e^2 N/m}$ is the plasma frequency; $e = -|e|$ and $m$ are the electron charge and mass, respectively; $N$ is the free electron density; $\varepsilon = \varepsilon_r + i\varepsilon_i$ is the complex dielectric permittivity of the bulk metal and can be represented from Drude model as $\varepsilon = \varepsilon_\infty - \omega_p^2/[\omega^2 + i\omega\nu(\omega)]$. The parameter $\varepsilon_\infty$ determined by polarizability of bound (core) electrons in metal and the frequency dependence $\nu(\omega)$ of the inverse relaxation time (effective collision frequency) should be chosen to provide best fit between functions $\varepsilon_{i,r}(\omega)$ and related measurement data for a particular material.

Employing Eq. (1) (along with boundary condition of zero normal current density at plasma boundary) allows one to generalize the known solution by Mie [31, 32] for the problem on collective electromagnetic oscillations of metal sphere and results in the following dispersion relation for the complex eigenfrequencies $\tilde{\omega} = \omega - i\gamma$ of electric multipoles of metal cluster in free space (see also [4, 5, 33]):

$$\varepsilon - \varepsilon_\infty = \frac{\varepsilon_\infty k_p a j_l'(k_p a)}{l(l+1)j_l(k_p a)}\left\{\frac{\varepsilon[k_0 a h_l(k_0 a)]'}{h_l(k_0 a)} - \frac{[k_t a j_l(k_t a)]'}{j_l(k_t a)}\right\}. \tag{2}$$

Here, $l = 0,1,2,3,\ldots$ is the multipole order; $h_l(\xi) = j_l(\xi) + in_l(\xi)$; $j_l(\xi)$ and $n_l(\xi)$ are the spherical Bessel functions of the first and second kind, respectively; the primes denote derivatives with respect to their arguments; $k_0 = \tilde{\omega}/c$ is the free-space wavenumber; $k_t = \tilde{\omega}\sqrt{\varepsilon}/c$ and $k_p = \sqrt{5\varepsilon/[3\varepsilon(\varepsilon_\infty - \varepsilon)]}/r_F$ are respectively the wavenumbers of transverse and longitudinal waves in degenerate cluster plasma.

In the quasistatic limit ($k_0 a \to 0$), Eq. (2) goes into the simpler equation [4, 11, 15]

$$\varepsilon_\infty \frac{\varepsilon l + l + 1}{\varepsilon_\infty - \varepsilon} = \frac{l(l+1)}{k_p a}\frac{j_l(k_p a)}{j_l'(k_p a)} \tag{3}$$

that can also be found directly through the solution to the following boundary problem for electric potential $\varphi(\mathbf{r})\exp(-i\omega t)$:

$$\Delta\varphi(r>a)=0; \quad \varphi(r<a)=\varphi_t+\varphi_p; \Delta\varphi_t=0; \quad \Delta\varphi_p+k_p^2\varphi_p=0; \tag{4}$$

$$\varphi\big|_{r=a-0}=\varphi\big|_{r=a+0}, \quad \varepsilon_\infty\frac{\partial\varphi}{\partial r}\bigg|_{r=a-0}=\frac{\partial\varphi}{\partial r}\bigg|_{r=a+0}, \quad \frac{\partial(\varepsilon\varphi_t-\varphi)}{\partial r}\bigg|_{r=a-0}=0. \tag{5}$$

Here, $r$ is a distance from the center of the cluster; the last boundary condition in (5) corresponds to the zero normal current at plasma boundary. This problem can be also reformulated into equations for the potential and charge density $\rho$:

$$\varepsilon_\infty\Delta\varphi+4\pi\rho=0, \quad \Delta\rho+k_p^2\rho=0, \quad \rho(r>a)=0, \tag{6}$$

with the boundary conditions (5), the last of which takes the form

$$\frac{\partial}{\partial r}\left(\varphi+\frac{12\pi}{5}r_F^2\rho\right)\bigg|_{r=a-0}=0. \tag{7}$$

The eigenfunctions in this problem can be expressed in spherical coordinates $r$, $\vartheta$, $\phi$ as follows:

$$\varphi_{lmn}=C_{lmn}R_{ln}(r)Y_{lm}(\cos\vartheta,\phi); \quad R_{ln}(r<a)=-\left(\frac{r}{a}\right)^l+f_{ln}(r); \tag{8}$$

$$f_{ln}(r)=\frac{l\varepsilon_{ln}+l+1}{l+1}\frac{j_l(k_{ln}r)}{j_l(k_{ln}a)}; \quad R_{ln}(r>a)=\frac{l\varepsilon_{ln}}{l+1}\left(\frac{a}{r}\right)^{l+1}; \tag{9}$$

$$\rho_{lmn}(r<a)=C_{lmn}\frac{k_{ln}^2}{4\pi a^2}f_{ln}(r)Y_{lm}(\cos\vartheta,\phi); \tag{10}$$

Here, $C_{lmn}$ are the arbitrary constants; $Y_{lm}(\vartheta,\phi)$ are the spherical harmonics (i.e., the products of the associated Legendre polynomials $P_l^{(m)}(\cos\vartheta)$ and trigonometric functions of azimuthal angle $\phi$; $m$, $l$ are integers with $|m|\leq l$); $k_{ln}=k_p(\varepsilon_{ln})$; $\varepsilon_{ln}$ are the solutions to Eq. (3), $n=0,1,2,3,\ldots$. As it can be seen, the degeneracy of order $2l+1$ with respect to $m$ takes place for any $l>0$ and any $n$.

Within the parameter range $r_F\ll a\ll k_0^{-1}$, $\varepsilon_i\ll 1$ being of the most theoretical and applied interest, the solution (8)-(10) describes the quasistatic multipole oscillations of two types (surface and volume plasmons) which differ significantly in the spatial distribution of the charge density. The surface plasmons ($n=0$) have almost zero charge density in the whole cluster volume except for a thin near-surface layer of thickness $\delta r\sim r_F$. The real parts of their eigenfrequencies (the solutions of Eq. (3) with $\varepsilon_i\equiv 0$) are approximately defined (with the terms of the order $r_F/a$ and $(k_0 a)^2$ omitted) by the equations

$$\varepsilon_r(\omega_{l0})l+l+1=0, \quad \omega_{l0}^2=\omega_p^2\frac{l}{(l+1)\varepsilon_\infty+1}. \tag{11}$$

The volume plasmons ($n\geq 1$) are standing waves of the volume charge density (solutions of the second equation in (6) with real wavenumbers $k_p$); their real eigenfrequencies are located in the region $\omega>\omega_p/\sqrt{\varepsilon_\infty}$:

$$\omega_{ln}^2=\frac{\omega_p^2}{\varepsilon_\infty}\left(1+\frac{3r_F^2}{5a^2}\mu_{ln}^2\right), \tag{12}$$

where $\mu_{ln}\equiv k_p(\omega_{ln})a$ are the roots of the equation $j_l(\mu)/j_{l'}(\mu)=\mu$; at $n\gg l$, these roots are $\mu_{ln}\approx(\pi/2)(l+1+2n)$.

Thus, there exists one surface plasmon ($n=0$) with frequency $\omega_{l0} < \omega_p/\sqrt{\varepsilon_\infty}$ and a sequence of volume plasmons ($n=1,2,3,...$) for every $l$. Since the used hydrodynamic approach (which assumes weak spatial dispersion) is valid only for not too large wavenumbers $k_t, k_p \ll r_F^{-1}$, both of the plasmon sequences are limited by conditions $l, n \ll a/r_F$. When these conditions are broken, the plasmons are almost fully suppressed by the strong Landau damping. Note that the above upper estimation for number of surface plasmons agrees with formal calculation of this number [15] from Eq. (3) without the Landau damping taken into account.

As noted above, the damping of both surface and volume plasmons is generally determined by three types of losses: bulk (determined by the imaginary part of material dielectric permittivity), surface, and radiative ones. If the losses are not too strong (with damping constant $\gamma \ll \omega$), they are in fact additive, that is, the total linewidth for every plasmon is a sum of the three corresponding components calculated independently, without other types of losses being considered, $\gamma_{ln} = \gamma_{ln}^{(v)} + \gamma_{ln}^{(r)} + \gamma_{ln}^{(s)}$. The first two terms of this sum can be found directly from Eq. (2). Without radiative losses (i.e., in the limits $k_0 a \to 0$, $k_t a \to 0$), this equation goes into Eq. (3) that does not contain imaginary unit in the explicit form and determines real values of $\varepsilon_{ln}$ and $k_p^2(\varepsilon_{ln})$. With given complex function $\varepsilon(\tilde{\omega})$, these values correspond to complex eigenfrequencies $\tilde{\omega}_{ln}$ in complex plane of $\tilde{\omega}$. The real parts of these eigenfrequencies are defined by Eqs. (11) and (12), and all the imaginary parts are determined, within the Drude model used, by the effective collision frequency $\nu(\omega)$ in the same way for all the plasmons, $\gamma_{ln}^{(v)} = \nu(\omega_{ln})/2$.

The radiative damping constant $\gamma^{(r)}$ can be found just through the replacement of the Bessel functions in Eq. (2) with the first terms of their power series with respect to the small parameters $k_0 a$, $k_t a$. In the absence of bulk losses (at purely real $\varepsilon = \varepsilon_r$), this results in the following equations for damping constants of surface and volume plasmons, respectively:

$$\gamma_{l0}^{(r)} = \omega_{l0} \left(\frac{\omega_{l0} a}{c}\right)^{2l+1} \frac{l+1}{l(2l+1)[1 \cdot 3 \cdot ... \cdot (2l-1)]^2} \qquad (13)$$

$$\gamma_{ln}^{(r)} = \frac{9\mu_{ln}^2 \omega_p}{25} \left(\frac{r_F}{a}\right)^4 \left(\frac{\omega_p a}{c}\right)^{2l+1} \frac{l}{(l+1)[1 \cdot 3 \cdot ... \cdot (2l-1)]^2} \qquad (14)$$

To estimate the surface damping constant, we consider a simple one-dimensional model for the conversion of the electron oscillatory energy into the thermal energy during elastic electron reflection (at some time instant $t = t_c$) from plasma boundary ($x = 0$) treated as a rectangular potential barrier. The electron normal velocity components $V_x^{(\pm)}$ before ($t \leq t_c$, $V_x^{(+)} > 0$) and after ($t \geq t_c$, $V_x^{(-)} < 0$) the reflection from the boundary and the alternating electric field $E_x \cos \omega t$ (assumed homogeneous) near the boundary are related through equations $V_x^{(\pm)} = V_{0x}^{(\pm)} + \tilde{V}_x \sin \omega t$, where $V_{0x}^{(\pm)}$ is the drift velocity component, and $\tilde{V}_x = eE_x/m\omega$ is the amplitude of the oscillatory component. The reflection elasticity [$V_x^{(-)}(t_c) = -V_x^{(+)}(t_c)$] allows one to find relation between drift components of the velocity before and after the reflection and the averaged (over various reflection times $t_c$) increase of the electron kinetic energy $\Delta w_0 = \langle (m/2)[V_{0x}^{(-)2} - V_{0x}^{(+)2}] \rangle$ which is determined by these drift components:

$$V_{0x}^{(-)} = -(V_{ox}^{(+)} + 2\tilde{V}_x \sin \omega t_c), \qquad (15)$$

$$\Delta w_0 = m(\tilde{V})^2 = m(eE_x/m\omega)^2. \tag{16}$$

Strictly speaking, some correction coefficient $K \sim 1$ is required in Eq. (16) to take into account the inhomogeneity of $E_x$ field near the boundary and the smoothness of the boundary (i.e., the fact the potential barrier is non-rectangular). We find the total power $Q$ of surface losses (energy which plasmon looses in time unit through the electron-surface collision on the whole cluster boundary surface $S$) by equating the plasmon energy loss in a single electron-surface collision to $\Delta w_0$ (with correction coefficient) and calculating the Fermi-distribution-averaged electron flux density $q = 3NV_F/16$ (for electrons moving towards the boundary):

$$Q = \oiint_S \Delta w_0 q dS = K \frac{3e^2 N V_F}{16m\omega^2} \oiint_S E_n^2 dS, \tag{17}$$

where $E_n = -\partial\varphi/\partial n$ is the normal component of the electric field at the boundary. Based on Eq.(17), equations for the eigenoscillation total energy

$$W = \frac{1}{16\pi}\left[\left(\varepsilon_\infty + \frac{\omega_p^2}{\omega^2}\right)\iiint_{r<a}|\mathbf{E}|^2 dV + \iiint_{r>a}|\mathbf{E}|^2 dV\right], \tag{18}$$

and its decrease rate

$$\frac{\partial W}{\partial t} = -2\gamma^{(s)}W = -Q, \tag{19}$$

we find the equations for constants $\gamma^{(s)} = Q/2W$ of damping caused by electron-surface scattering for surface and volume plasmons, respectively,

$$\gamma_{l0}^{(s)} = \frac{3KV_F \oiint_S |\partial\varphi_{lm0}/\partial r|^2 dS}{16\iiint_{r<a}|\nabla\varphi_{lm0}|^2 dV} = \frac{3KV_F(l+1)^2}{16al}, \tag{20}$$

$$\gamma_{ln}^{(s)} = -\frac{3KV_F\omega_p^2 \oiint_S |\partial\varphi_{lmn}/\partial r|^2 dS}{8(\varepsilon_\infty\omega^2 + \omega_p^2)\iiint_{r<a}\varphi_{lmn}\Delta\varphi_{lmn}dV} = \frac{27K\varepsilon_\infty^2 l^2 \mu_{ln}^2}{200}\left(\frac{r_F}{a}\right)^4 \frac{V_F}{a}. \tag{21}$$

The obtained equation (20) for surface damping constant is in a good agreement with the known size dependence of the electron-surface scattering rate $\gamma^{(s)} \sim V_F/R$, which was previously obtained for the dipole surface plasmon from various theoretical models and experimental data [6, 29, 34-38] (and which can be in particular used to refine the value of correction coefficient $K$ for various cluster materials). Note that the present dependence of the damping constant on multipole order ($\gamma_{lm0}^{(s)} \sim (l+1)^2/l$) differs from similar dependence ($\gamma_{lm0}^{(s)} \sim l$) obtained previously in Ref. [36], where the surface plasmon field inside the cluster (including near-surface region) was found without spatial dispersion taken into account (with the use of the local dielectric function ).

As for the volume plasmons, their surface and radiative losses are much smaller than these ones of the surface plasmons (as it follows from Eqs. (13), (14), (20), and (21), $\gamma_{ln}^{(s)}/\gamma_{l0}^{(s)} \sim \gamma_{ln}^{(r)}/\gamma_{l0}^{(r)} \sim n^2 l(r_F/a)^4 <<1$) and the contribution of these two losses to the total damping constant occur to be very small as compared to the volume losses. This fact (which originates from smallness of the field amplitude at cluster boundary for $n \neq 0$) makes the accuracy of determining the factor $K$ in Eq. (21) for the volume plasmons not so important.

## III. CALCULATION OF FREQUENCY AND ENERGY SPECTRA OF PLASMONS EXCITED BY A MOVING ELECTRON

The problem on excitation of collective electron oscillations in a cluster by a source having arbitrary spatiotemporal distribution of the external charge density $\rho_s(\mathbf{r},t)$ can be solved by the same expansion method which is used for problem on excitation of any distributed oscillatory system by the external source. By applying the time-domain description to Eq. (1) (i.e., with substituting $-i\omega \to \partial/\partial t$) and based on equations relating the electrical displacement and field strength to the external ($\rho_s$) and induced ($\rho$) charge densities,

$$\nabla \mathbf{D} = 4\pi \rho_s, \quad \varepsilon_\infty \nabla \mathbf{E} = 4\pi(\rho + \rho_s) \text{ at } r < a, \tag{22}$$

$$\nabla \mathbf{E} = 4\pi \rho_s \text{ at } r > a, \tag{23}$$

we obtain the following equations which rule the spatiotemporal evolution of the induced charge density (at $r < a$) and the potential within the quasistatic potential approach ($\mathbf{E} = -\nabla \varphi$) for arbitrary function $\rho_s(\mathbf{r},t)$,

$$\frac{\partial^2 \rho}{\partial t^2} + 2\hat{\gamma}\frac{\partial \rho}{\partial t} + \frac{\omega_p^2}{\varepsilon_\infty}\rho - \frac{3}{5}V_F^2 \Delta \rho = -\frac{\omega_p^2}{\varepsilon_\infty}\rho_s, \tag{24}$$

$$\varepsilon_\infty \Delta \varphi = -4\pi(\rho + \rho_s) \text{ at } r < a; \tag{25}$$

$$\Delta \varphi = -4\pi \rho_s \text{ at } r > a. \tag{26}$$

The boundary conditions for this problem are as follows:

$$\varphi\big|_{r=a-0} = \varphi\big|_{r=a+0}, \quad \varepsilon_\infty \frac{\partial \varphi}{\partial r}\bigg|_{r=a-0} = \frac{\partial \varphi}{\partial r}\bigg|_{r=a+0}, \quad \frac{\partial}{\partial r}\left(\varphi + \frac{12\pi}{5}r_F^2 \rho\right)\bigg|_{r=a-0} = 0. \tag{27}$$

In equation (24), $\hat{\gamma}$ is the damping operator incorporating all the mentioned mechanisms of losses. Let $\rho_{st}(\mathbf{r},t)$ and $\varphi_{st}(\mathbf{r},t)$ be the solution of the corresponding static problem (without terms with time derivatives in Eq. (23))

$$\frac{\omega_p^2}{\varepsilon_\infty}\rho_{st} - \frac{3}{5}V_F^2 \Delta \rho_{st} = -\frac{\omega_p^2}{\varepsilon_\infty}\rho_s, \tag{28}$$

$$\varepsilon_\infty \Delta \varphi_{st} = -4\pi(\rho_{st} + \rho_s) \text{ at } r < a; \tag{29}$$

$$\Delta \varphi_{st} = -4\pi \rho_s \text{ at } r > a$$

with the same boundary conditions (27). Then the differences $\rho_{pl}(\mathbf{r},t) = \rho(\mathbf{r},t) - \rho_{st}(\mathbf{r},t)$ and $\varphi_{pl}(\mathbf{r},t) = \varphi(\mathbf{r},t) - \varphi_{st}(\mathbf{r},t)$ present the dynamical (truly plasmonic) part of the solution and satisfy (at $r < a$) the equations

$$\frac{\partial^2 \rho_{pl}}{\partial t^2} + 2\hat{\gamma}\frac{\partial \rho_{pl}}{\partial t} + \frac{\omega_p^2}{\varepsilon_\infty}\rho_{pl} - \frac{3}{5}V_F^2 \Delta \rho_{pl} = \frac{\partial^2 \rho_{st}}{\partial t^2} + 2\hat{\gamma}\frac{\partial \rho_{st}}{\partial t}, \tag{30}$$

$$\varepsilon_\infty \Delta \varphi_{pl} = -4\pi \rho_{pl}, \tag{31}$$

They can be expanded in eigenfunctions (8)-(10) of the boundary problem (4), (5) (or the equivalent problem (6), (7)):

$$\rho_{pl} = \sum_\alpha G_\alpha(t)\rho_\alpha(\mathbf{r}), \quad \varphi_{pl} = \sum_\alpha G_\alpha(t)\varphi_\alpha(\mathbf{r}). \tag{32}$$

Here and later, the set of indices $l, m, n$ is replaced by a single index $\alpha$ or $\beta$ for the sake of brevity. The orthogonality relation required to find such expansion can be obtained from Eqs. (6) and boundary condition (7) (see Appendix) and is written as follows:

$$\iiint\limits_{r<a} \rho_\alpha \tilde{\varphi}_\beta \, dV = 0 \text{ at } \alpha \neq \beta, \qquad (33)$$

where $\tilde{\varphi}_\beta = \varphi_\beta + 12\pi r_D^2 \rho_\beta/5$. Through the standard procedure for finding time-dependent factors $G_\alpha(t)$ (substituting Eq. (32) into Eqs. (24)-(25), multiplying the result by $\tilde{\varphi}_\beta$, and integrating over the nanoparticle volume, we obtain

$$\frac{d^2 G_\alpha}{dt^2} + 2\hat{\gamma}_\alpha \frac{dG_\alpha}{dt} + \omega_\alpha^2 G_\alpha = \frac{1}{N_\alpha}\left( \frac{d^2 F_\alpha}{dt^2} + 2\hat{\gamma}_\alpha \frac{dF_\alpha}{dt} \right), \qquad (34)$$

where $F_\alpha = \iiint\limits_{r<a} \rho_{st}(\mathbf{r},t)\tilde{\varphi}_\alpha(\mathbf{r})dV = \iiint\limits_\infty \rho_s(\mathbf{r},t)\varphi_\alpha(\mathbf{r})dV$, and $N_\alpha = \iiint\limits_{r<a} \rho_\alpha \tilde{\varphi}_\alpha dV$ is the mode norm. These equations clearly demonstrate the nature of the temporal evolution of any mode as that one of a linear oscillator excited by the given external force.

In the case we are interested in, the external charge is an electron moving with a constant velocity $V_0$ in a straight line which is separated from the cluster center by a distance $b$, and the electron position is determined by equations $x = b$, $y = 0$, $z = V_0 t$, $-\infty < t < \infty$ in the Cartesian coordinates $x$, $y$, $z$. In these coordinates, we have

$$\rho_s = e\delta(x-b)\delta(y)\delta(z-V_0 t), \qquad (35)$$

$$F_\alpha = e\varphi_\alpha(b, 0, V_0 t), \qquad (36)$$

i.e., the temporal behavior of the external force $F_\alpha(t)$ exciting a particular mode is fully determined by the eigenfunctions (8), (9) profile at the electron trajectory (with the following arguments of the spherical functions: $r = \sqrt{b^2 + (V_0 t)^2}$, $\cos\vartheta = V_0 t/r$, and $\phi = 0$).

The total energy which electron looses due to its deceleration by the field of the induced charges inside the cluster is defined by the plasmonic part of the solution $\varphi_{pl}$ (the contribution from the static part $\varphi_{st}$ is zero) and can be calculated in the both time-domain and frequency-domain representations:

$$W = \sum_\alpha eV_0 \int_{-\infty}^{+\infty} G_\alpha(t) E_\alpha(t) \, dt = \sum_\alpha eV_0 \int_{-\infty}^{+\infty} G_{\alpha\omega} E_{\alpha\omega}^* \, d\omega. \qquad (37)$$

Here $E_\alpha = -\partial\varphi_\alpha(b,0,z)/\partial z\big|_{z=V_0 t}$ is the $z$ component of the induced electric field at electron position, and the Fourier spectrum of this component is

$$E_{\alpha\omega} = \frac{i\omega}{2\pi V_0^2} \Phi_\alpha(b, \frac{\omega}{V_0}) \qquad (38)$$

$$\Phi_\alpha(b, \frac{\omega}{V_0}) = \int_{-\infty}^\infty \varphi_\alpha(b,0,z)\exp(i\omega z/V_0)dz \qquad (39)$$

As it follows from Eqs. (34) and (39), the Fourier spectrum of $G_\alpha(t)$ is

$$G_{\alpha\omega} = -\frac{e(\omega^2 + 2i\gamma_\alpha\omega)}{2\pi N_\alpha V_0(\omega_\alpha^2 - \omega^2 - 2i\gamma_\alpha\omega)} \Phi_\alpha(b,0,z). \qquad (40)$$

Since $\varphi_\alpha$ is purely real and the product $\Phi_\alpha \Phi_\alpha^*$ is an even function of frequency $\omega$, the Eq. (37) can be rewritten as $W = \int_0^\infty W_\omega d\omega$ with

$$W_\omega = \frac{e^2}{\pi V_0^2} \sum_\alpha \frac{2\omega^2 \omega_\alpha^2 \gamma_\alpha}{(\omega_\alpha^2 - \omega^2)^2 + 4\gamma_\alpha^2 \omega^2} \frac{1}{N_\alpha} \left| \Phi_\alpha \left( b, \frac{\omega}{V_0} \right) \right|^2. \tag{41}$$

On the other hand, using the known quantum-mechanical correspondence between frequency- and energy-dependent loss spectra, i.e., treating $W$ as the quantum average of some energy loss probability distribution for many electrons incident on the cluster [18] ($W = \int_0^\infty \hbar\omega \, (dP/d\omega) \, d\omega$; $\hbar$ is the reduced Planck constant), we obtain the equation relating the loss frequency spectrum (41) and the differential probability of loosing a particular energy value $\hbar\omega$ for a fixed value of the impact parameter $b$,

$$\frac{dP}{d\omega} = \frac{W_\omega}{\hbar\omega}. \tag{42}$$

By integrating this quantity over the infinite transverse plane $z = \text{const}$, we find the total differential probability of energy loss for the cluster bombarded by a broad electron beam,

$$\frac{dP_\Sigma}{d\omega} = 2\pi \int_0^\infty \frac{dP}{d\omega} b \, db. \tag{43}$$

## IV. NUMERICAL RESULTS AND DISCUSSION

Below we present the computer simulation results illustrating the important features of the energy loss spectra, which have been observed in previous experiments, but have not been interpreted within theoretical and numerical models providing the clear enough physical understanding and reliable identification of these features. As it follows from the equations obtained, the contribution to the loss spectrum from excitation of any particular oscillation mode is a product of the imaginary part $\text{Im}[(\omega^2 - \omega_\alpha^2 + 2i\omega\gamma_\alpha)]^{-1}$ of the Lorentz-shaped resonant response and the excitation factor

$$U_\alpha = \frac{1}{N_\alpha} \left| \Phi_\alpha \left( b, \frac{\omega}{V_0} \right) \right|^2 \tag{44}$$

which depends (with $a$, $b$, and $\omega/V_0$ fixed) on the spatial structure of the excited mode. The ratio between excitation factors of various modes depends significantly on the position of the electron trajectory with respect to the cluster, which fact underlies the changes in spectrum shape with impact parameters in scanning transmission electron microscopy (STEM) experiments employing thin (compared to the cluster radius) electron beam (see, e.g., Refs. [20-23]). These changes are illustrated in Fig. 1 for the silver cluster of radius 12 nm, which is bombarded by 300-keV electrons ($V_0 = 2.3 \times 10^{10}$ cm/s). The loss spectra calculated from Eqs. (41) and (42) and obtained experimentally [21] are shown in Figs 1(a) and 1(b), respectively, for the same sequence of the impact parameter values increasing from bottom to top curves with the step 2 nm (from $b = 0$ to $b = 10$ nm). The calculations use the parameters of Ag $\varepsilon_\infty = 3.83$, $\omega_p = 9.1 \, \text{eV}$ taken from Ref. [39]; effective electron collision frequency $\nu(\omega)$ was found from curves $\varepsilon_r(\omega)$ and $\varepsilon_i(\omega)$ presented in Ref. [40]. The correction coefficient $K$ in Eq. (20) for surface damping constant is chosen to be 1/3 on the basis of results of Ref. [35].

The first (at lower energy $\hbar\omega \approx 3.2 \pm 0.1 \, \text{eV}$) of two peaks in spectra in Fig. 1 originates from excitation of dipole ($l = 1$) and quadrupole ($l = 2$) surface plasmons, which are merged together owing to the strong enough damping. The relative contribution of a particular plasmon in the total loss changes with the impact parameter due to the change of the excitation factor, and this can be traced by the shift of the total loss probability maximum towards lower energies, from the quadrupole resonance $\hbar\omega = 3.3\,\text{eV}$ (in the bottom curve with $b = 0$) to dipole resonance $\hbar\omega = 3.1\,\text{eV}$ (in the top curve with $b = 10\,\text{nm}$).

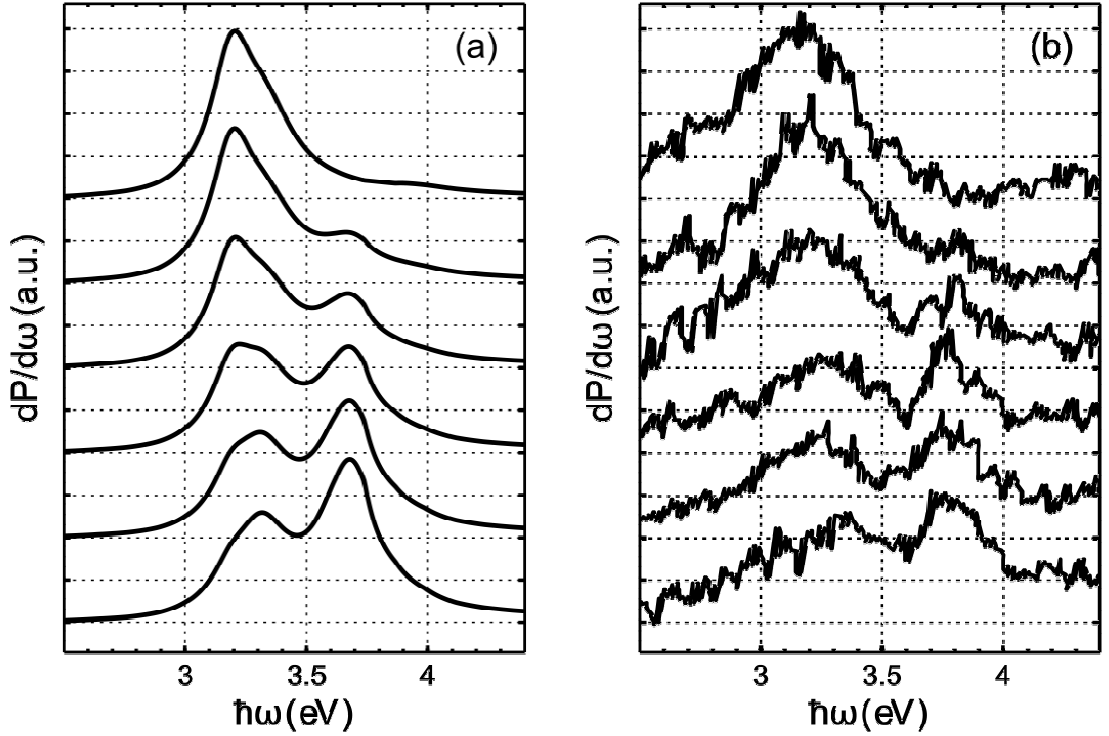

FIG. 1. (a) Theoretical and (b) experimental energy loss spectra for Ag cluster (radius $a = 12\,\text{nm}$) irradiated by 300-keV electrons with different impact parameters $b$. In both panels, different curves (from bottom to top) correspond to $b = 0, 2, 4, 6, 8, 10\,\text{nm}$. Theoretical curves are obtained with Eqs. (34), (35); and the experimental results are taken from [21]. In calculations, the parameters of Ag were chosen to fit the data in Refs. [39, 40].

The second peak (at energy $\omega = 3.8\,\text{eV}$ corresponding approximately to the condition $\varepsilon_r(\omega) = 0$) originates from excitation of the lowest (with $n = 1$) spherically symmetric ($l = 0$) and quadrupole ($l = 2$) volume plasmons. These resonance are also merged due to proximity of the plasmon frequencies. The contribution to total loss from other volume plasmons is negligible owing to fast decrease of the excitation factor with $l$ and $n$. As it can be seen from plotted curves, the volume plasmon can be effectively excited only if electron travels through the central region of the cluster, when the corresponding peak in the energy loss probability exceeds peak associated with surface plasmons. With increasing the impact parameter, the excitation efficiency decreases for the volume plasmons while for the surface plasmons (mostly for the dipole one) it grows, reaching a maximum at grazing incidence ($b \approx a$). The comparison of calculated and experimentally measured loss spectra in Fig. 1 brings out good enough agreement between them, which may indicate that the general approach used and the choose of models and parameters for damping constant calculation are adequate.

The number of multipole modes contributing noticeably to the total loss spectrum should obviously depend significantly on the dimensionless passage time $\tau = \omega_p a/V_0$, i.e., on the ratio between the passage time of electron through the interaction region and the characteristic time scale $\omega_p^{-1}$ of the collective oscillations [11-13]. With that parameters being small, the spectrum is determined only by the lowest modes ($l = 0,1,2$) as in the above example (with $\tau = 0.2$). With $\tau$ increased, the higher modes start playing a role. In particular, as it can be seen from Fig. 2 where the loss spectra are shown for Al cluster with different number of modes taken into the account, the addition of new modes stops changing the total spectrum only from $l, n \sim 7-10$ at $\tau = 2$ already.

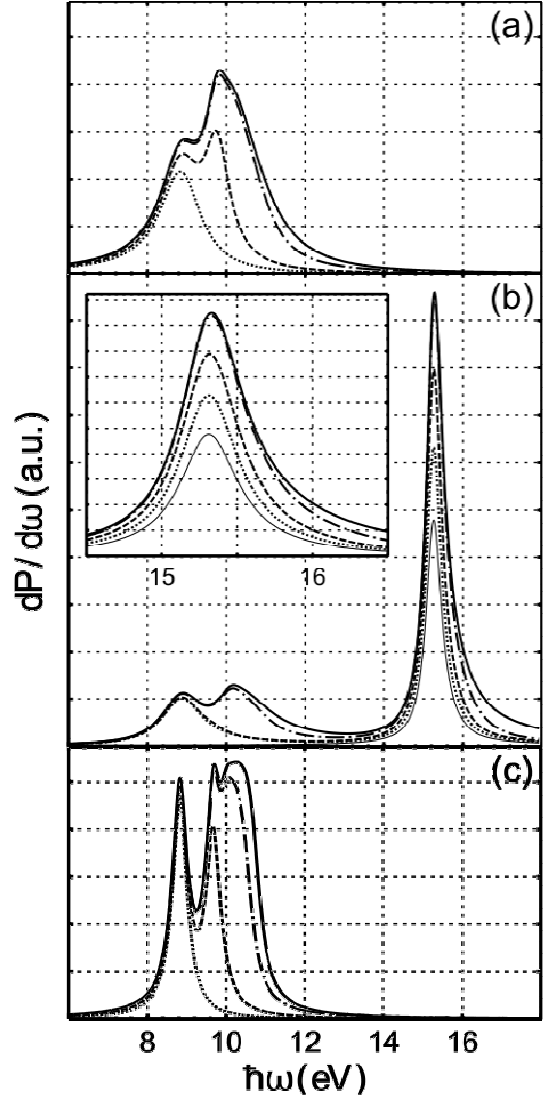

FIG. 2. Energy loss spectra of Al cluster demonstrating the contribution of different multipole modes for the cases of grazing incidence (a, c) and central one (b), correspondingly; radius of cluster $a = 10$ nm, electron energy 50 keV. Panel (c) shows energy loss spectra calculated in Ref. [12] without radiation and surface damping taken into account. In each panel, thin solid line presents contribution of the symmetric modes ($l = 0$); the dotted line does so for symmetric and dipole modes ($l = 0,1$); the dashed and dash-dotted lines incorporate additionally modes with $l \leq 2$ and $l \leq 10$, correspondingly; the solid lines show the overall energy loss spectra.

This applies to the both grazing [$b = a$, Fig. 2(a)] and central [$b = 0$, Fig. 2(b)] incidence, i.e., to the excitation of both surface and volume plasmons. To illustrate the role of various dissipation mechanisms, we also plot the results of similar calculation performed previously in Ref. [12] for grazing incidence without radiation and surface damping (and without spatial dispersion in cluster plasma) taken into account, see Fig. 2(c). As it is seen, for the given cluster size ($a = 10$ nm), the radiative damping (see also Ref. [41]) strongly suppresses the dipole surface plasmon resonance [$\hbar\omega = 8.8$ eV, Fig. 2(a)] while this resonance is pronounced without radiative damping [Fig. 2(c)] as strong as the quadrupole one ($\hbar\omega = 9.7$ eV) is. As for the surface damping, it results in faster convergence of the sum (34) and broadening of the right part in the aggregated resonance peak. Note that dipole and quadrupole resonance are not merged in this case unlike in the above loss spectrum for Ag cluster. This is due to the lesser value of parameter $\varepsilon_\infty$ (close to 1 for Al), which determines the frequency intervals between multipoles of different orders as it follows from Eqs. (11).

Figure 3 demonstrates the changes in the energy loss spectrum shape with the dimensionless passage time $\tau$ from 1 to 4 (the higher values are hardly possible if the electron mean free path in a cluster is greater as compared to cluster radius). The calculation are performed for free-electron cluster model with $\varepsilon_\infty = 1$, $\omega_p = 2\times10^{16}$ s$^{-1}$, $\nu/\omega_p = 0.02$; electron velocity $V_0 = 10^{10}$ cm/s. The mentioned range of $\tau$ corresponds to the cluster radius range from $a = 5$ nm to $20$ nm. The figure presents the probability distributions $dP/d\omega$ at the impact parameters $b = a$ [Fig. 3(a)] and $b = 0$ [Fig. 3(b)] as well as loss spectra $dP_\Sigma/d\omega$ averaged over electrons with various impact parameters in the broad electron beam [Fig. 3(c)]. All the spectra are normalized to their maximum values. In the surface plasmon frequency band ($\omega_{0l}/\omega_p = 1/\sqrt{2+1/l}$, $l = 1,2,3,\ldots$), the averaged spectra are of similar shapes to those in the case of grazing incidence [$b = a$, Fig. 3(a)], where the excitation efficiency of surface plasmon occurs to be at maximum for any dimensionless passage time $\tau$. At $\tau \sim 1$ (as well as at $\tau \ll 1$), the dipole plasmon ($l = 1$, $\omega/\omega_p \approx 0.58$) dominates among the excited surface plasmons, but with $\tau$ increasing, it gradually gives way to the quadrupole plasmon ($\omega/\omega_p \approx 0.65$) at first and then to the multipoles of higher order whose resonance peaks are merged together (due to their convergence and growth of the surface damping constant) and form one broad maximum at frequency $\omega_{max}$ slightly lower than limit frequency $\omega_\infty$ for surface plasmons, $\omega_\infty/\omega_p = 1/\sqrt{2} \approx 0.7$. One can find the order $l_{max}(\omega_{max})$ of the corresponding multipole, for large $\tau$ by estimating the spatial range $\Delta z$ and multipole order range where the approximate phase synchronism $V_0 - \omega_{0l} a/l < \omega_{0l}\Delta z$ takes place between the spatial harmonics of the multipole field and electron moving with velocity $V_0$ at $b = a$. At $\tau \gg 1$, such estimation gives $l_{max} \approx \tau/\sqrt{2}$, which is in a qualitative agreement with curves in Figs. 3(a) and 3(c) at large $\tau$.

The volume plasmons, due to small intervals between them, are presented in the loss spectrum by a single peak of width $\nu/2$ with a maximum at $\omega_p$ in all cases. For central incidence, this peak is much higher than any peak corresponding to surface plasmon [Fig3(b)]. After averaging over impact parameter both peaks are of the same height. [Fig3(c)].

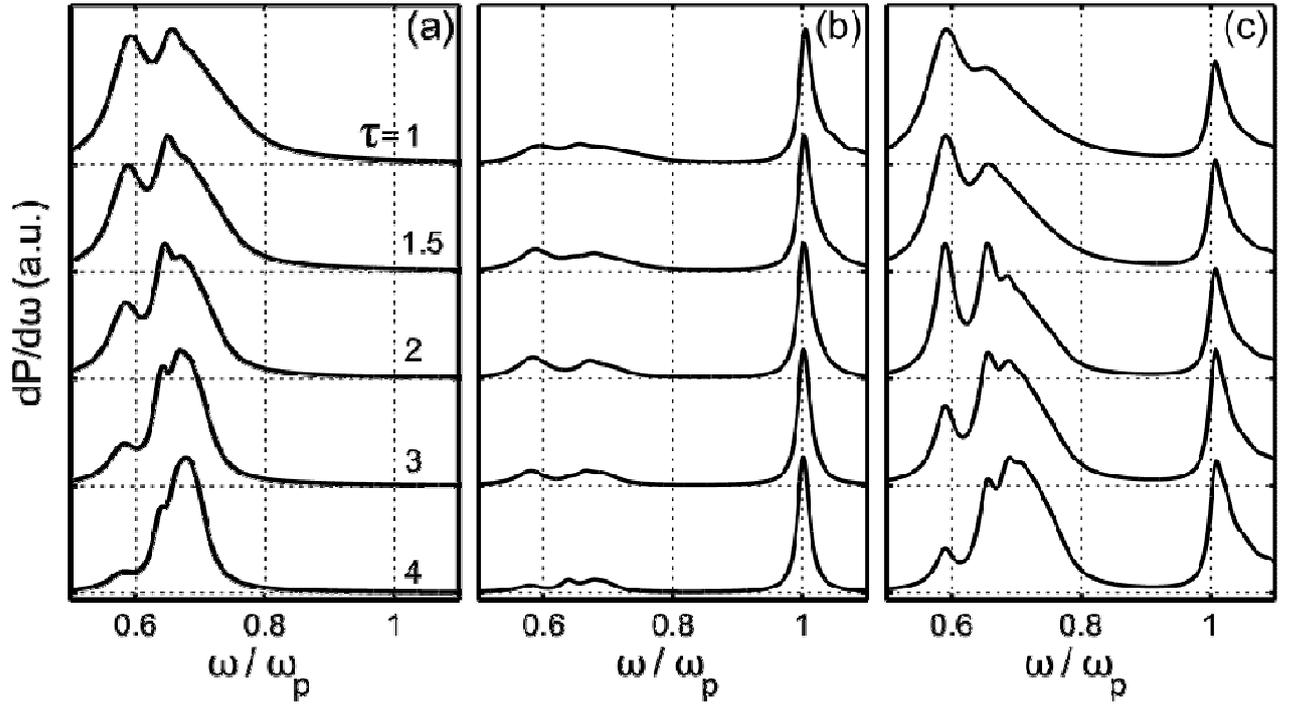

FIG. 3. (a) Energy loss spectra for different values of the dimensionless passage time $\tau$ in cases of (a) grazing incidence (the impact parameter $b = a$), (b) central one ($b = 0$), and (c) broad electron beam (with averaging over impact parameter). The cluster parameters are $\omega_p = 2 \times 10^{16}$ s$^{-1}$, $\nu/\omega_p = 0.02$; electron velocity is $V_0 = 10^{10}$ cm/s. All the spectra are normalized to their maximum values.

## V. CONCLUSION

We calculated the spectra of multipole eigenoscillations of the electric types excited in a spherical metal cluster by an electron passing near or through it in the framework of the hydrodynamic approach. When calculating the damping constants of the plasmons, we allowed for three main types of energy losses, namely, volume losses determined by the imaginary part of the complex dielectric function of the cluster material, surface losses caused by the reflection of the free electrons in the cluster from the cluster boundaries, and radiation losses caused by electromagnetic radiation of the corresponding multipole. In the context of the problems of inelastic electron scattering from a metal cluster, the second and third loss mechanisms have not yet been paid sufficient attention in previous investigations. As it follows from the formulas we obtained for the damping constants determined by these mechanisms, this lack of attention can be justified only for the volume plasmons, for which the surface and radiation losses are negligibly small compared to the volume (dielectric) losses. As for the surface plasmons, all three types of losses turn to be of comparable significance in that region of cluster radius values ($\sim 5-30$ nm) where the plasmon resonances are strongest. Radiation losses dominate for the dipole plasmon at the upper boundary of this region [for the $l$-order multipole the radiation damping constant $\gamma_l^{(r)} \sim \omega(\omega a/c)^{2l+1}$]. Approaching to the lower boundary results in the growth of the surface losses. The corresponding damping constant, which is determined by the energy conversion rate from ordered oscillatory electron motion to the thermal one during electron reflection from the cluster boundary, increases with a decrease in the cluster radius and an increase in the multipole order [$\gamma_l^{(s)} \sim (V_F/a)(l+1)^2/l$], in contrast to the radiation damping constant.

We found general analytical formulas for the excitation coefficients of surface and volume cluster plasmons in the time domain and frequency domain representations, and calculated numerically the energy loss spectra of moving electron (differential probabilities for the loss of any given energy value) using the calculated damping constants for plasmons of different types. The calculated variations with impact parameter of the type of loss spectra agree well with the data of the STEM experiments and have allowed us to describe the character of the loss spectrum evolution (both for narrow and wide electron beams) with change of the electron passage time $\tau = \omega_p a / V_0$.

## AKNOWLEDGMENTS

The work (V. B. G. and I. A. P.) was supported by the Government of the Russian Federation (Agreement No. 14.B25.31.0008) and the Russian Foundation for Basic Research (Grants No. 13-02-00964, 14-02-00847, and 14-02-31722). The solution of the excitation problem in the part concerning the orthogonality relation and the preparation of Appendix (V. A. K.) was supported by the Russian Science Foundation (Grant No. 15-12-10033).

## APPENDIX: PROOF OF THE ORTHOGONALITY RELATION

Here, we obtain the orthogonality relation that is required for solution of Eqs. (23) and (24) through expansion in plasmonic modes. For that purpose, we write down Green's second identity for functions $\tilde{\varphi}_{\alpha,\beta} = \varphi_{\alpha,\beta} + (12\pi/5) r_F^2 \rho_{\alpha,\beta}$ inside the cluster,

$$\iiint_{r<a} \left( \tilde{\varphi}_\alpha \Delta \varphi_\beta - \tilde{\varphi}_\beta \Delta \varphi_\alpha \right) dV = \oiint_{r=a} \left( \tilde{\varphi}_\alpha \frac{\partial \varphi_\beta}{\partial n} - \tilde{\varphi}_\beta \frac{\partial \varphi_\alpha}{\partial n} \right) dS, \tag{A1}$$

and expand the Laplace operators and normal derivatives applied to $\tilde{\varphi}_{\alpha,\beta}$ in this identity using Eqs. (6) and boundary condition (7). After that, we have

$$\iiint_{r<a} \left[ \left(1 + \frac{3k_{p\alpha}^2 r_F^2}{5}\right) \rho_\alpha \tilde{\varphi}_\beta - \left(1 + \frac{3k_{p\beta}^2 r_F^2}{5}\right) \rho_\beta \tilde{\varphi}_\alpha \right] dV = 0. \tag{A2}$$

Here $k_{p\alpha,\beta}$ are the wavenumbers of longitudinal waves in degenerate cluster plasma for corresponding modes. Since the charge densities $\rho_{\alpha,\beta}$ are absent outside the cluster, the integration domain in the last equation can be extended to the whole infinite space. The reciprocity theorem $\iiint_\infty \rho_\alpha \tilde{\varphi}_\beta dV = \iiint_\infty \rho_\beta \tilde{\varphi}_\alpha dV$ allows one to simplify (A2) to $(k_{p\alpha}^2 - k_{p\beta}^2) \iiint_\infty \rho_\alpha \tilde{\varphi}_\beta dV = 0$, and therefore at $k_{p\alpha}^2 \neq k_{p\beta}^2$ we have

$$\iiint_{r<a} \rho_\alpha \tilde{\varphi}_\beta = 0 \quad \text{at} \quad \alpha \neq \beta. \tag{A3}$$

## REFERENCES

1. J.W.M. Chon, and K. Iniewski, *Nanoplasmonics: Advanced Device Applications. Devices, Circuits, and Systems* (Boca Raton: CRC Press, 2013).
2. V.V. Klimov, *Nanoplasmonics* (Boca Raton: CRC Press, 2014).

3. *Plasmonics: from Basics to Advanced Topics. Springer Series in Optical Sciences,*. Vol. v.167, ed. by S. Enoch, and B. Nicolas. (Berlin: Springer 2012).
4. V.B. Gildenburg, I.G. Kondrat'ev, Radio Eng. Electr. Phys. **10**, 560 (1965).
5. R. Ruppin, Phys. Rev. B, **11**, 2875 (1975).
6. V. B. Gildenburg, V. A. Kostin, and I. A. Pavlichenko, Phys. Plasmas **18**, 092101 (2011).
7. J. Hurst, F. Haas, G. Manfredi, P.-A. Hervieux, Phys. Rev. B. **89**, 161111(R) (2014).
8. M. Kuisma, A. Sakko, T.P. Rossi, A.H. Larsen, J. Enkovaara, L. Lehtovaara, and T.T. Rantala, Phys. Rev. B **91**, 115431 (2015).
9. V.P. Krainov, B.M. Smirnov, M.B. Smirnov, Physics-Uspekhi **50**, 907 (2007).
10. R.A. Ganeev, *Laser-Surface Interactions* (Dordrecht: Springer Netherlands, 2014).
11. H.F. Fujimoto and K. Komaki, J. Phys. Soc. Jpn **25**, 1679 (1968).
12. T.L. Ferrell, P. M. Echenique, Phys. Rev. Lett. **55**, 1526 (1985).
13. T.L. Ferrell, R.J. Warmack, V.E. Anderson, and P.M. Echenique, Phys. Rev. B **35**, 7365 (1987).
14. N. Barberan and J. Bausells, Phys. Rev. B **31**, 6354 (1985).
15. M.T. Michalewicz, Phys. Rev. B **45**, 13664 (1992).
16. L.G. Gerchinkov, A.V. Solov'yov, J.-P. Connerade, W. Greiner, J. Phys. B: At. Mol. Opt. Phys. **30**, 4133 (1997).
17. L.G. Gerchikov, A.N. Ipatov, R.G. Polozkov, A.V. Solov'yov, Phys. Rev. A **62**, 043201 (2000).
18. F. J. García de Abajo, Rev. Mod. Phys, **82**, 209, (2010).
19. L. Kiewidt, M. Karamehmedović, C. Matyssek, W. Hergert, L. Mädler, T. Wriedt, Ultramicroscopy, **133**, 101 (2013).
20. M. Acheche, C. Colliex, H. Kohl, A. Nourtier, P. Trebbia, Ultramicroscopy, **20**, 99 (1986).
21. A.L. Koh, et al., ACS Nano **3**, 3015 (2009).
22. J.A. Scholl, A.L. Koh, J.A. Dionne, Nature **483**, 421 (2012).
23. S.Raza, S Kadkhodazadeh, T. Christensen, M. D. Vece, M. Wubs, N. A. Mortensen, and N. Stenger, arXiv:1505.00594v1 [physics.optics] (2015)
24. M. Rocca, Surf. Sci. Rep. **22**, 1-71 (1995)
25. R.F. Egerton, *Electron Energy-Loss Spectroscopy in the Electron Microscope* (New York: Springer, 2011).
26. T. Hanrath and B.A. Korgel, Nano Lett. **4**, 1455 (2004).
27. J. Nelayah, M. Kociak, O. Stephan, F. Javier, G. de Abajo, et al. Nature Phys. **3**, 348 (2007).
28. M. Picher, S. Mazzucco, S. Blankenship, R. Sharma, Ultramicroscopy, **150**, 10 (2015).
29. S. Holst, W. Legler. Z. Phyz. D: At. Mol. Clus. **25**, 261 (1993).
30. F. Haas, *Quantum Plasmas: an Hydrodynamic Approach* (New York: Springer, 2011).
31. G. Mie, Ann. Phys. Leipz. **25**, 377 (1908)
32. J. A. Stratton, *Electromagnetic theory*. (McGraw-Hill Book Company, New York, 1941).
33. A. Yildiz, Nuovo Cimento **30**, 1182 (1963).
34. H. Hövel, S. Fritz, A. Hilger, U. Kreibig, and M. Vollmer, Phys. Rev. B, **48**, 18178 (1993).
35. U. Kreibig, M. Gartz, and A. Hilger, Ber. Bunsen-Ges. Phys. Chem. **101**, 1593 (1997).
36. C. Yannouleas, R. A. Broglia. Ann. Phys. **217**, 105 (1992).
37. B.N.J. Persson, Surf. Sci. **281**, 153 (1993).
38. S. Berciaud, L. Cognet, Ph. Tamarat, and B. Lounis, Nano Lett. **5**, 515 (2005).
39. H. Gai, J. Wang, and Q. Tian, Appl. Opt. **46**, 2229 (2007).
40. P. B. Johnson and R. W. Christy Phys. Rev. B **6**, 4370 (1972).
41. F. J. Garcia de Abajo, Phys. Rev. B **59**, 3095 (1999).